\documentclass[3p]{elsarticle}
\usepackage[T1,T2A]{fontenc}
\usepackage[utf8]{inputenc}
\usepackage[english]{babel}
\usepackage{lineno,hyperref}
\usepackage{graphicx}
\usepackage{subfig}
\usepackage[dvipsnames]{xcolor}
\usepackage[]{amsmath, amssymb} 
\usepackage{marginnote}
\modulolinenumbers[5]
\usepackage{soul}
\usepackage{marginnote}
\usepackage{bookmark}
\DeclareUnicodeCharacter{2212}{-}
\DeclareUnicodeCharacter{03C3}{-}
\DeclareUnicodeCharacter{03BC}{-}

\newcommand{\beginsupplement}{%
\setcounter{table}{0}
\renewcommand{\thetable}{S\arabic{table}}%
\setcounter{figure}{0}
\renewcommand{\thefigure}{S\arabic{figure}}%
}

\journal{}

\newcommand{\Eon}{E_{\mathrm{ON}}}
\newcommand{\Eoff}{E_{\mathrm{OFF}}}
\newcommand{\EMF}{\eta}

\begin{document}

\begin{frontmatter}
 
\title{Ultrafast opto-mechanical terahertz modulators based on stretchable carbon nanotube thin films}


\author[mymainaddress]{Maksim I. Paukov}

\author[mysecondaddress]{Vladimir V. Starchenko}
\author[myfourthaddress]{Dmitry V. Krasnikov}

\author[mysixthaddress]{Gennady A. Komandin}

\author[myfourthaddress]{Yuriy. G. Gladush}

\author[mymainaddress]{Sergey S. Zhukov}
\author[mymainaddress]{Boris P. Gorshunov}
\author[myfourthaddress]{Albert G. Nasibulin}
\author[mymainaddress]{Aleksey V. Arsenin}
\author[mymainaddress]{Valentyn S. Volkov}

\author[mymainaddress,myfivethaddress]{Maria G. Burdanova}

\address[mymainaddress]{Center for Photonics and 2D Materials, Moscow Institute of Physics and Technology, 9 Institutskiy ln, Dolgoprudny, 141701 Moscow region, Russia}
\address[mysecondaddress]{Moscow Institute of Physics and Technology, 9 Institutskiy ln, Dolgoprudny, 141701 Moscow region, Russia}

\address[myfourthaddress]{Skolkovo Institute of Science and Technology, 3 Nobel St., Moscow, 121205 Russia}

\address[mysixthaddress]{Prokhorov General Physics Institute of the Russian Academy of Sciences, 38 Vavilov St., Moscow, 119991 Russia}

\address[myfivethaddress]{Institute of Solid State Physics of the Russian Academy of Sciences, 2 Academician Osipyan St., Chernogolovka, 142432 Moscow region, Russia}

\begin{abstract}
For terahertz (THz) wave applications, tunable and rapid modulation is highly required.  When studied by means of optical pump-terahertz probe spectroscopy, single-walled carbon nanotubes (SWCNTs) thin films demonstrated ultrafast carrier recombination lifetimes with a high relative change in the signal under optical excitation, making them promising candidates for high-speed modulators. Here, combination of SWCNT thin films and stretchable substrates facilitated studies of the SWCNT mechanical properties under strain, and enabled the development of a new type of an opto-mechanical modulator. By applying a certain strain to the SWCNT films, the effective sheet conductance and therefore modulation depth can be fine-tuned to optimize the designed modulator. Modulators exhibited a photoconductivity change of 3-4 orders of magnitude under the strain due to the structural modification in the SWCNT network. Stretching was used to control the THz signal with a modulation depth of around 100\% without strain and 65\% at a high strain operation of 40\%. The sensitivity of modulators to beam polarisation is also shown, which might also come in handy for the design of a stretchable polariser. Our results give a fundamental grounding for the design of high-sensitivity stretchable devices based on SWCNT films.

\end{abstract}


\end{frontmatter}

\section{Introduction}
The application of terahertz (THz) radiation has been found in many spheres including explosive and concealed weapon detection\cite{Federici2005}, pharmacy\cite{ZEITLER2009} etc. Since the beginning of the century THz waves have also been considered as a means of wireless data transfer, which possesses much better operation properties than current gigahertz (GHz) and microwave technologies\cite{2007}. Several significant advantages of THz communication are mentioned in \cite{Federici2010}: increased bandwidth capacity, small diffraction of waves in free space, ability to provide secure communication, a small atmospheric scintillation effect, and low attenuation of THz radiation under weather conditions\cite{Sarieddeen2020}. THz modulation is a critical factor, which also has to be considered in developing ultrafast THz 6G-technologies, in addition to generation, propagation, and detection\cite{Federici2010}.

The amplitude of THz waves passed through a medium is determined by its conductivity. Therefore, the electrical, optical or thermal\cite{Rahm2012} manipulation of the conductivity provides a way to control the amplitude of THz radiation. The quality of typical modulators is characterized by several parameters, among which are modulation depth (MD) and modulation speed (MS). It has been reported that traditional bulk semiconductors, such as GaAs, Ge and Si, can have either high MD and low MS (in order of several ns) \cite{Rahm2012}, or fast MS (in order of several ps) and insufficient MD ($<$40\%) under the optical excitation\cite{Beard2001}. Compared to 3D counterparts, optically-controlled low-dimensional materials (LDMs), like graphene, transition metal dichalcogenides (TMDs), carbon nanotubes (CNT) and nanowires, have demonstrated superior MD up to 99\% and MS up to 600 GHz\cite{Chen2018}. Electrically-tuned modulators, based on LDMs, such as GaAs/AlGaAs and GaN/AlGaN nanoheterostructures\cite{Kersting2000},\cite{KleineOstmann2004}, monolayer graphene\cite{Chen2018}, also possess giant MD factor of 93-99\%.

Meanwhile, stretchable THz modulators with high efficiency are rarely observed. MD of modulators based on 2D materials was rather small due to limited change in conductivity under applied mechanical deformation\cite{Cheng2018,Liu2016}. This problem can be overcome using SWCNT films, which have shown a great potential in optoelectronics\cite{Kharlamova2022,Kharlamova2022a}, especially in stretchable devices\cite{Gilshteyn2019}. In addition, CNTs have also been proposed as one of the ideal candidates for THz modulation combining both high MS and MD. In \cite{BURDANOVA2021}, we showed that these fruitful results are due to the negative THz photoconductivity effect under optical excitation, which is widely studied nowadays\cite{KAR2022100631}. Some practical results for THz applications, obtained by constructing stretchable CNT modulators, have been reported\cite{Xu2018}.

In this work, we report the study on the influence of stretching on optically-controlled modulators based on SWCNT. The designed modulators exhibit photoconductivity change of 3-4 orders under 40\% stretching, which can be used to control the THz optical modulator with giant initial MD close to 100\%. To evaluate the possible explanation for the MD change under the applied strain, we discuss the possible mechanisms of photoconductivity change such as partial alignment, consequent formation of bundles\cite{Gubarev2019} and formation of microcracks. The proposed tunable modulator paves a path for devices capable of switching the desirable properties of the THz signal.

\section{Materials and methods}

SWCNTs were synthesized by aerosol CVD method based on on the carbon monoxide disproportionation (the Boudouard reaction) on the surface of Fe-based catalyst\cite{Khabushev2019}, \cite{Khabushev2020}. To prepare a thin film, The filtration of SWCNT aerosol with a nitrocellulose membrane was followed by the dry transfer on the transparent stretchable substrate: 0.2 mm thick elastomer (Silpuran, Wacker)\cite{Kaskela2010}. Two types of SWCNT films with thicknesses of approximately 11 and 106 nm (correspond to optical transmittance of 90\% and 60\% at 550 nm wavelength, respectively) were obtained. All the thin films consisted of nearly equilibrium compositions: one-third metallic and two-thirds semiconducting SWCNTs\cite{Nasibulin2005}. 

SWCNT film on the elastomer is placed between two jaws of the home-made mechanical stretcher with the ability to control elongation precisely. The film and its holder were placed horizontally and vertically into the spectrometer so that the probe THz beam polarisation was perpendicular and parallel according to the stretching direction. 


The pump beam for the optical pump, THz probe (OPTP) spectroscopy was created by an optical parametric amplifier (TOPAS), seeded by a 1\,kHz, 40\,fs, 800\,nm pulse (from a Newport Spectra Physics Spitfire ACE) to create pulses with a tunable center wavelength. Here, a pump wavelength of 600\,nm was used (above the $E_{\mathrm{11}}$ and $E_{\mathrm{22}}$ excitonic absorption lines of the CNTs). As the used SWCNTs have narrow bandgaps, any wavelength shorter than 2.5\,$\mu$m can be applied. The probe beam was generated using home-made conventional THz-TDS setup as described in \cite{Monti2018} allowing us to measure both equilibrium conductivity and photoconductivity. The extraction of the conductivity and photoconductivity were performed according to our previous study\cite{Burdanova2019}.

\section{Concept and results}

\begin{figure}[tb]
\begin{center}
 \subfloat{\includegraphics[width=1\textwidth]{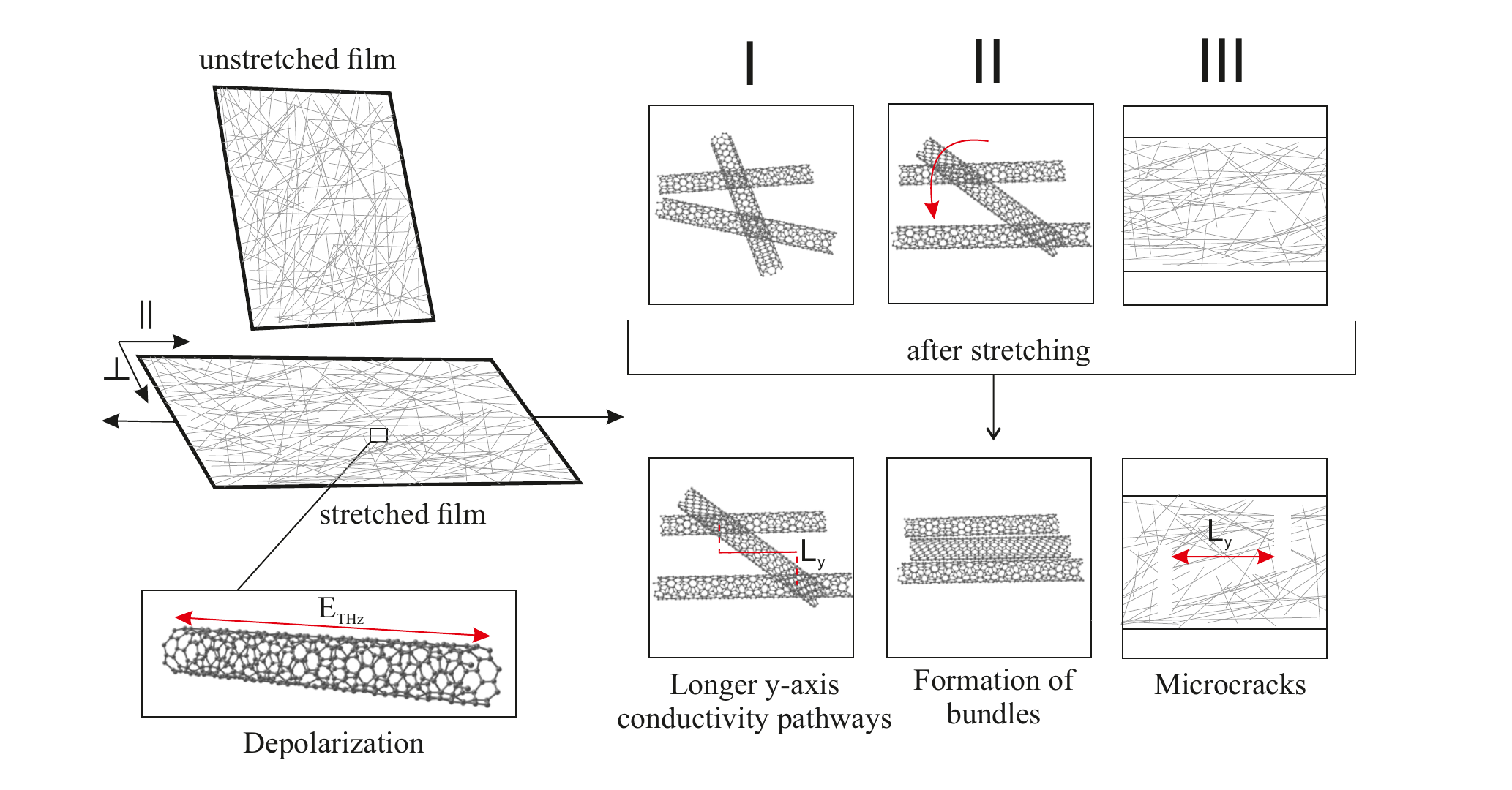}}
\end{center}
\caption{\label{FIG:Concept} Illustration of typical changes of SWCNT film morphology formed by stretching. The stretching results in: I. junction movement and partial alignment of SWCNT under strain; II. Formation of bundles due to the close location of aligning SWCNT; III. formation of microcracks under applied stresses. Two typical experiments were performed - perpendicular and parallel to stretching and following compression directions. The partial alignment results in anisotropic properties coming from the depolarisation effect.}
\end{figure}

The morphology of the SWCNT network changes under strain\cite{Gilshteyn2019}, which leads to the modification of optoelectronic properties (Fig. \ref{FIG:Concept}\marginpar{f1}). Therefore, we would like to describe firstly the possible changes in the film that occur over the stretching. Stretching in one direction results in the alignment of SWCNT\cite{Gilshteyn2019}. This consequently alters carrier conduction pathways making it longer in the direction parallel to stretching due to the movement of SWCNT junctions and shorter in the perpendicular direction. Here and further we use a term "conduction pathways" to describe both the interaction between free charges and the change of the percolation of the CNT network.   To evaluate the anisotropy towards THz radiation, we have calculated the factor $Re( \sigma_{\mathrm{\perp}})/Re (\sigma_{\mathrm{\parallel}})$,where $Re (\sigma_{\mathrm{\perp}})$, and $Re (\sigma_{\mathrm{\parallel}})$ are the values of photoconductivity, which were obtained at the frequency of 1 THz for different relative elongations. The results are presented in Fig. \ref{FIG: SI Anisotropy} of SI. The reduction of the perpendicular component in comparison to the parallel component of the real part of the conductivity is present in the stretching direction, while the opposite behavior was observed in the compression direction. This indicates the gradual alignment and cracks formation process which irreversibly modify the effective conductivity of the CNT network. 

The further increase of stretching results in bundling primarily due to the reduced spacing between the SWCNTs. It has been reported that the conductivity of the SWCNT bundle is lower for bigger diameter bundles of SWCNTs\cite{Shuba2012}.
The significant stretching results in the formation of microcracks. A strong relationship between resistance and these crack lengths has already been shown in DC-measurements in \cite{Shindo2012}. The significant resistance increase was observed there due to the shortening of conductivity pathways in the direction parallel to stretching. In the same direction cracks on individual SWCNT appeared due to the applied stresses. These factors also limit conductivity pathways but on a nanometer scale. Finally, the partial alignment causes the difference in the conductivity and photoconductivity in parallel and perpendicular directions in reference to the THz polarisation beam due to the depolarisation effect, according to which optical absorption is suppressed when the polarisation of the incident radiation is perpendicular to the SWCNT axis\cite{Islam2004}.

\begin{figure}[tb]
\begin{center}
 \subfloat{\includegraphics[width=1\textwidth]{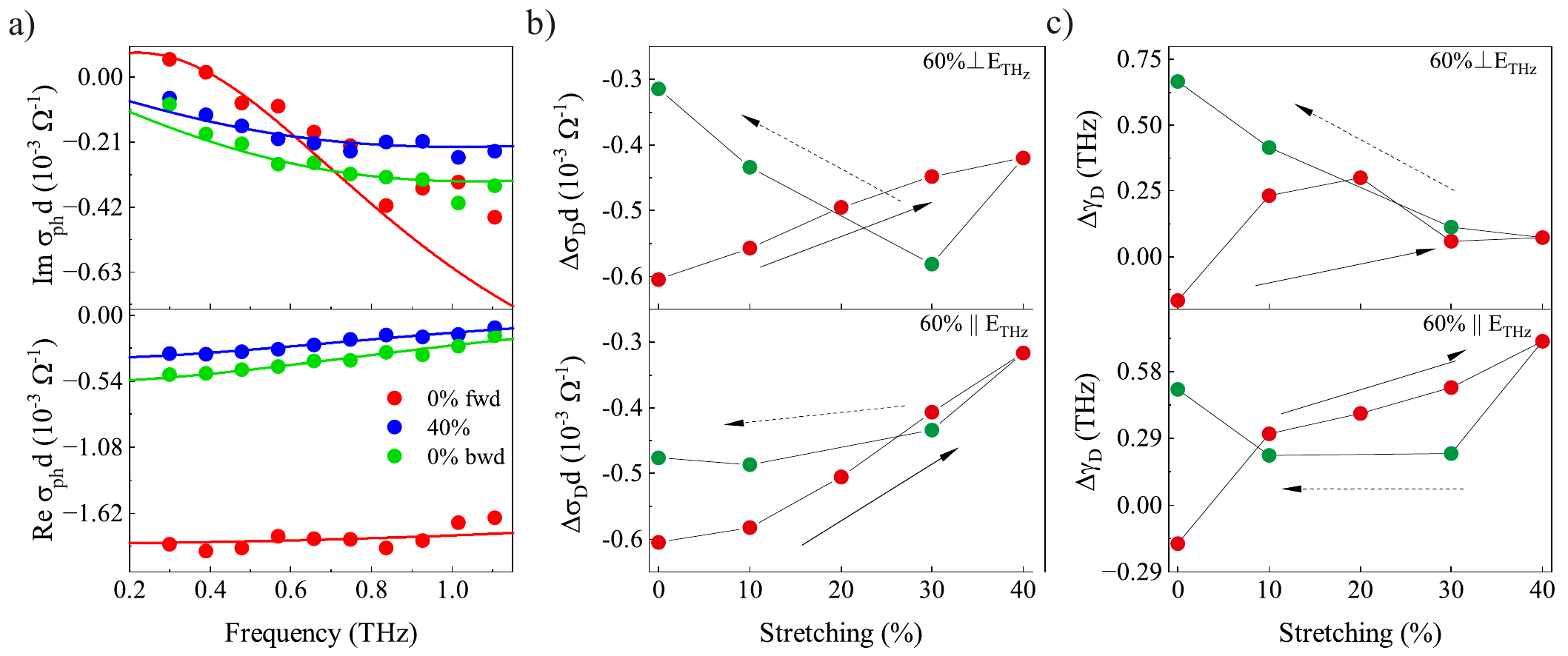}}
\end{center}
\caption{\label{FIG: photoconductivity} (a) The film sheet photoconductivity spectra at 3 ps after photoexcitation of unstretched SWCNT film of 60\% transparency ($\epsilon$=0\%, red color), stretched film ($\epsilon$=40 \%, blue color) and film after compression ($\epsilon$=0\%, green color). The change of the DC conductivity (b) and scattering rate (c) obtained from fitting with the change of the Drude model.}
\end{figure}

In our previous studies, \cite{Burdanova2019,BURDANOVA2021}, we showed that thin SWCNT films exhibit giant negative THz photoconductivity associated with the trion formation under optical excitation, which results in the uniquely high MD and MS. In the study, presented in this paper, we carried out stretching and following compressing experiments on the same samples with stretching up to 40\,\%. Rather than reporting the local, microscopic response functions of each SWCNT, we presented the effective conductivity which averages the response over the percolated CNT network. Before being stretched, films showed typical for SWCNT photoconductivity as reported in \cite{BURDANOVA2021}. The photoconductivity spectra were obtained at $t$ = 3 ps pump-probe
delay time by using the relation described in \cite{Burdanova2021b}. As shown on Fig. \ref{FIG: photoconductivity}a\marginnote{f2a}, the absolute value of the real part of THz photoconductivity decreases monotonically with the increasing of the strain in both perpendicular and parallel direction to the probe beam polarisation. Photoconductivity changes over stretching and, moreover, is an irreversible feature, which will be indicated further in other characteristics. Full data for all stretching parameters are available in Fig. \ref{FIG: SI Spectra}, \ref{FIG: SI Spectra1} SI.

To reveal the nature of change in photoconductivity in the case of the experiments with stretching, we fitted the conductivity of the samples under optical excitation $\sigma_{\mathrm{ex}}(\omega)$ and without it $\sigma_{\mathrm{eq}}(\omega)$ with the Drude-Lorentz model for each value of stretching. This model accounts two key physical processes, which affect THz conductivity: the confined collective motion of carriers along nanotubes (plasmon contribution) and free carrier response, indicative of delocalized intertube
transport along percolating channels on a larger scale (Drude contribution)\cite{Zhukova2017}, \cite{Gorshunov2018}. The conductivity in the case of the Drude-Lorentz model reads as follows:
\begin{equation}
  \sigma(\omega)=\sigma_{\mathrm{D}}\frac{i\gamma_{\mathrm{D}}}{\omega+i\gamma_{\mathrm{D}}}+\sigma_{\mathrm{p}}\frac{i\omega\gamma_{\mathrm{p}}}{\omega^{\mathrm{2}}-\omega_{0}^{\mathrm{2}}+i\omega\gamma_{\mathrm{p}}},
\end{equation}

where $\sigma_D$ is the DC-conductivity of the Drude part, $\sigma_p$ is the plasmon conductivity at the resonance frequency $\omega_0$ (Lorentz part),  $\gamma_D$, $\gamma_p$ are scattering rates for free-electron and plasmon response, respectively. To understand the spectral dependence, we fitted photoconductivity $\sigma_{\mathrm{ph}}(\omega, t)$ =$ \sigma_{\mathrm{ex}}(\omega) − \sigma_{\mathrm{eq}}(\omega)$. We assumed that the plasmonic contribution
to the conductivity was unchanged by photoexcitation following our trion interpretation\cite{Burdanova2019}. Indeed, the negative photoconductivity in our experiments can be modeled as a drop in momentum scattering rate accompanied by lower DC conductivity for the Drude model. This indicates that after photoexcitation either the free carriers density is lowered or the average effective mass of the formed quasiparticles is increased\cite{Burdanova2019}.

We report then the influence of stretching on the change in charge carrier transport. As it was mentioned earlier, we considered only the difference between the Drude components of the conductivity in photo-excited state and equilibrium one. The changes in the fitting parameters $\Delta\sigma_{\mathrm{D}}=\sigma_{\mathrm{D}}^{\mathrm{ex}}-\sigma_{\mathrm{D}}^{\mathrm{eq}}$ and $\Delta\gamma_{\mathrm{D}}=\gamma_{\mathrm{D}}^{\mathrm{ex}}-\gamma_{\mathrm{D}}^{\mathrm{eq}}$ in relation to the relative elongation are presented in Fig. \ref{FIG: photoconductivity}b and c\marginnote{f2bc}. At 0\% of stretching, under the optical excitation the lower change in the DC conductivity accompanied by the reduction of scattering rate was observed. This is consistent with our trion interpretation\cite{Burdanova2019}. The further increase in stretching results in the monotonical decrease of the change in DC conductivity and a higher scattering rate in both parallel and perpendicular directions of the stretching. In terms of free charge behavior, it can be regarded as the lower charge carrier density of free charges under the photoexcitation with a simultaneous decrease in the time between the free charge collisions. This might be interpreted as the decline in the number of the photoinduced charge carriers at high stretching and simultaneous reduction of carriers' mean free path. Moreover, less number of free charges were observed in the perpendicular direction in comparison to the parallel. This is consistent with the possible mechanisms presented in Fig. \ref{FIG:Concept}. Indeed, the formation of microcracks limits the conductivity pathways resulting in the increase of scattering probability. Also, exciton trapping and quenching at cracks result in fewer trion formation and therefore less free charges in the system and lower photoconductivity. 

The stretching process promotes parallel partial alignment of SWCNT. Due to the anisotropic properties, nanotubes aligned in the same direction as the pump pulse's polarisation are preferentially photoexcited. The resulted exciton and hence incident THz radiation with the same polarisation will be absorbed by these carriers. Stretching also might affect the film in a way that parallel to the direction of elongation conductive pathways become longer while perpendicular shorten due to the movement of SWCNT on the contacts and partial alignment. 

\begin{figure}[tb]
\begin{center}
 \subfloat{\includegraphics[width=1\textwidth]{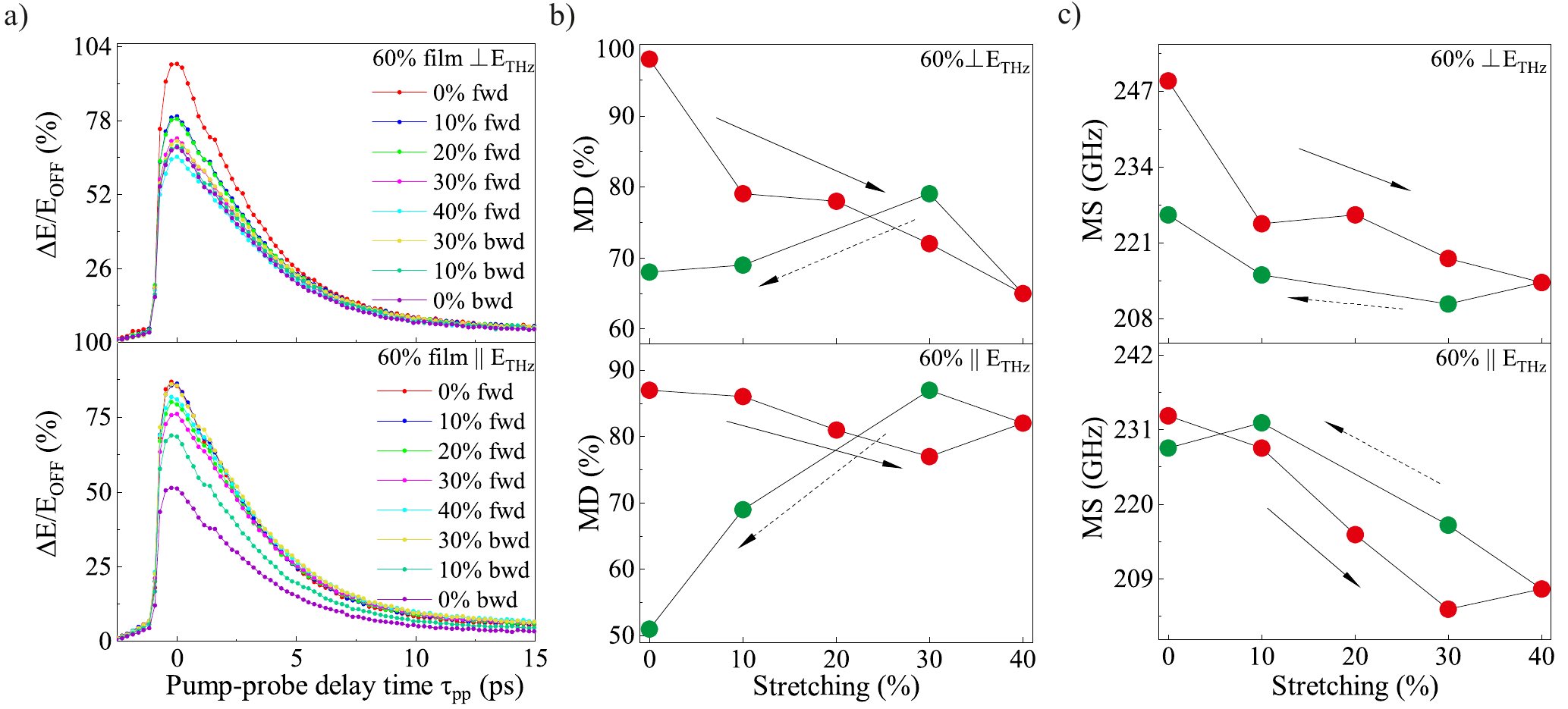}}
\end{center}
\caption{\label{FIG: Transient} (a) Modulation depth in transmission as a function of pump-probe delay time when the elongation is perpendicular (top) and parallel (bottom) to the THz probe beam. In transmission mode MD (b) and MS (c) for 60\% sample at stretching varying from 0 to 40\% and compression from 40 to 0\%. The arrows indicate the direction of stretching and the following compression.}
\end{figure}

Another possible explanation for the change of photoconductivity is the movement of SWCNT on contacts. As the strain increases, the SWCNT and film as a whole experience considerable shrinkage in the parallel direction and the SWCNT tend to align through sliding. At a certain level of strain, the overall distance between the SWCNTs becomes small enough to form bundles\cite{Liu2014},\cite{Nam2016}. CNTs have a tendency to form bundles which results in lowering conductivity and fewer number of free charges available in the system\cite{Shuba2012}. This also leads to a fewer trions, appearing in the system, and therefore, lower photoconductivity. At large stretching, the bundles are separated into several microblocks with relatively small deformations within these blocks. Once this deformation takes place, the characteristics are irreversible and diffident trends of photoconductivity is observed in two directions. In addition, sliding of SWCNTs results in a change of the photoconductivity at SWCNT intertube junctions due to the change of the distance between CNTs. Therefore, the change in intertube junctions also results in more packed networks and a change in the photoconductivity. 

A more complicated picture was observed in compression experiments. While in the parallel direction the relative recovery of the fitting parameters was observed, in the perpendicular direction the further decrease in the photoconductivity and a decrease of the scattering rate were found. This significant difference in the parallel and perpendicular stretching direction suggests that different processes acquire during the recovery. In particular, the conductivity difference along and perpendicular to the stretching direction should be caused by different scattering rate. Also, the change in the scattering rate is higher in the perpendicular direction. This indicates the change in the conductivity pathways, which might be explained as the formation of microcracks along the SWCNT and film. Microcracks therefore lead to the scattering from a boundary layer energy barrier, which also modify the charge carrier transport in parallel direction. In addition, in the compression experiments these microcracks still exist, working as a wire grid polariser, resulting in higher transparency in parallel direction and lower in perpendicular. Moreover, the depolarisation effect generates lower absorption in a perpendicular direction which consequently provides lower photoconductivity. 

We also would like to point out features at around 0.7\,THz for the sample with 90\% transparency that were not accounted for by the fitting model, but which has been previously reported\cite{KAR2019731,Xu2009,Lui2014,Burdanova2019,PhysRevMaterials.3.026003}, and which may be linked to intra-excitonic transitions (for more details see Fig. \ref{FIG: SI Spectra2}, \ref{FIG: SI Spectra3} in SI). Importantly, the peak position moves towards higher frequency as the result of stretching. This behaviour can be attributed to the influence of defects in electronic band structures and therefore in intra-excitonic transitions. 

In this section, we show that the stretching of the film leads to the emergence of competing mechanisms of changes in photoconductivity due to morphological modifications, which are connected with the formation of the CNTs texture and the growth of the number of microdefects.

\if The fundamentals of charge carrier transport under the light pump are studied by means of the analysis of measured THz field in frequency domain. First, the frequency dependence of the electric field is recalculated into the dependence of effective conductivity of the set of nanotubes as a whole on frequency, using Kramers Kroenig relation between the conductivity and the electric permittivity: (FORMULA). It is performed for both cases of non-illuminated and illuminated films. The data on conductivity of the former $\sigma$d are fitted then with Drude Lorentz model, which accounts two key factors, affecting THz conductivity: the movement of free charges (Drude contribution) and bounded ones (plasmon, or Lorentz, contribution). This model reads as follows: , where. The conductivity in the case of illumination $\sigma$1d is fitted with the same parameters of a plasmon, but with changed Drude contribultion. This interpretation is determined by. \fi

\if The extracted photoconductivity data СИГМАph d (omega) = СИГМА1d (omega) СИГМАd (omega) is fitted with Drude model (delta cигма D, delta гамма D) and presented on the Fig It is seen from that СИГМАph d (omega) dramatically changes over stretching and, moreover, is an irreversible feature, which will be also indicated further in other characteristics.\fi

\section{Ultrafast optical modulation}

\begin{figure}[t!]
\begin{center}
 \subfloat{\includegraphics[width=1\textwidth]{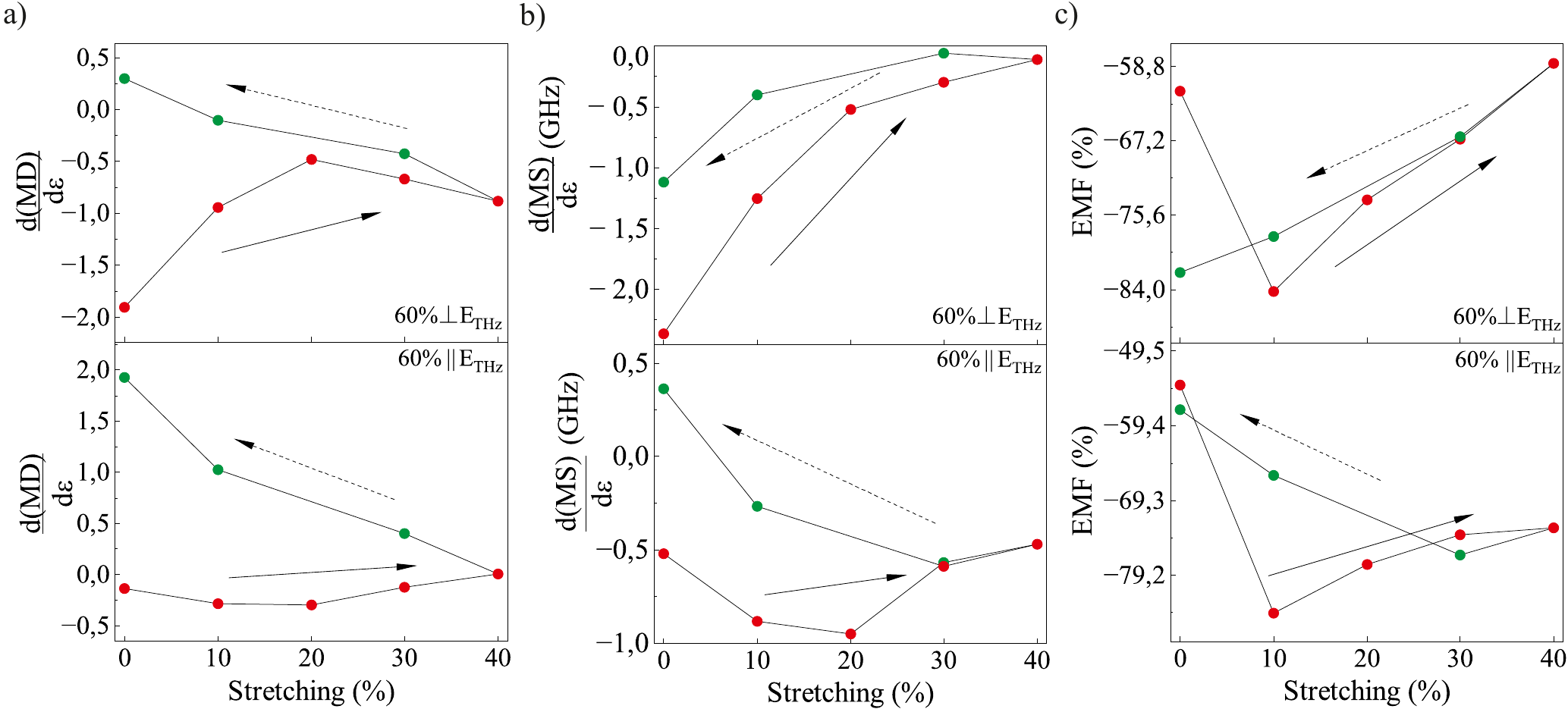}}
\end{center}
\caption{\label{FIG: OPTP-THz} Strain sensitivity of MD (a), MS (b) and EMF (c) as a function of stretching varying from 0 to 40\% and compression from 40 to 0\%. The arrows indicate the direction of stretching and following compression.}
\end{figure}

The transient phenomena, obtained by varying the time between photoexcitation by optical pump and THz probe beam allows for designing a THz-signal modulator with certain parameters. These modulators are typically characterized by several parameters, among which are modulation depth (MD) and modulation speed (MS). 
Modulation depth (MD) is defined as follows: $MD = \Delta E/\Eoff=(\Eon-\Eoff)/(\Eoff)$, where $\Eon, \Eoff$ are the detected THz signals after the sample with and without optical pump. MD in our experiments depended on several parameters: time delay, frequency, stretching, pump fluence.  To further explore the influence of stretching on ultrafast THz modulators, we measured MD as a function of stretching, which varied the optical density in the visible and THz ranges simultaneously. MD of the SWCNT at the peak of the THz pulse, which corresponds to a frequency-averaged response as a function of time after photoexcitation in perpendicular and parallel stretching direction towards the polarisation of the THz beam is shown in Fig. 3a.\marginnote{f3a} As it can be seen SWCNT film with the transparency of 60\% exhibits high MD in the range of nearly 50--100\%. With an increase in stretching the MD decreases monotonically from approximately 88 to 55\% in parallel direction and from 100 to 70\% in perpendicular direction. The modulator reached its highest MD within a few ps, before reverting back to its stationary level after approximately 15 ps. Here we defined the maximum operation frequency as MS = 1/$\tau$ using exponential fits to $\Delta E/E$ with lifetime of the photoinduced charge carriers $\tau$. Modulators, based on SWCNT of 60\% transparency, show high modulation speed of 205-250 GHz, while MS for a thinner film (with the transparency of 90\%, accordingly) varies approximately from 160 to 205 GHz. MS is also irreversible, depending on the orientation, and tends to go backward only halfway from non-stretching to 40\% stretched.

In order to understand the dependence of MD and MS on stretching, we plotted the smoothed derivative of both parameters (Fig. \ref{FIG: OPTP-THz}\marginnote{f4}). Despite the absence of the tendency in the curves, it is seen that the sharpest change in parameters occurs at the very beginning of stretching in most cases. The derivative of the MS increases monotonically with the stretching and slightly recovers in the compression in a perpendicular direction. The maximum value of -2.5 for the MS derivative at 0\% means that, the MS of the THz wave changes by -2.5\% for per change of strain 1\%, while -1 after compression at 0\% indicates the -1\% change per change of strain 1\%. This indicates the deterioration of MS strain sensitivity in the perpendicular direction. Different behavior was observed in parallel direction: MS almost recovered after stretching. The derivative for MD shows a value of -2 which also indicates a good initial sensitivity to strain which then switches to 0.5. However, in the parallel direction, it increases from almost 0 to 2.

One more quality of a modulator that was calculated is the energy modulation factor (EMF). It is given by the following formula:

\begin{equation}
 \EMF = \frac{\int \left | \Eoff(\omega) \right |^{\mathrm{2}}d\omega - \int \left | \Eon(\omega ) \right |^{\mathrm{2}}d\omega}{\int \left | \Eoff(\omega ) \right |^{\mathrm{2}}d\omega},
\end{equation}

\noindent where the integration is performed over the range of THz pulse frequencies. This metric is a way to evaluate the amount of energy of transmitted THz beam when the sample is optically excited compared to the equilibrium case. For normal
modulators (based on photoinduced absorption) $\eta$ is positive, as the energy of the transmitted beam is lower under photoexcitation. SWCNT films with transparency of 60\% demonstrate again $\eta$ varying -84 to -59\%. This indicates that such modulators exhibit high modulation parameters even under significant applied stresses. 

As a stretching device, SWCNT thin films can be also characterized by stretching modulation factor defined as follows:

\begin{equation}
SMF = \frac{MD(0\%)-MD(\epsilon=0...40\%)}{MD(0\%)},
\end{equation}

\noindent where $MD(0\%)$ is the modulation depth of a non-stretched film. When the elongation reaches 40\%, SMF is about 35\% in the perpendicular direction and 11\% in the parallel direction. At the end of the stretching cycle, it reaches values of 30\% and 45\% in perpendicular and parallel directions respectively.

\section{Discussion}

\begin{figure}[t!]
\begin{center}
 \subfloat{\includegraphics[width=0.75\textwidth]{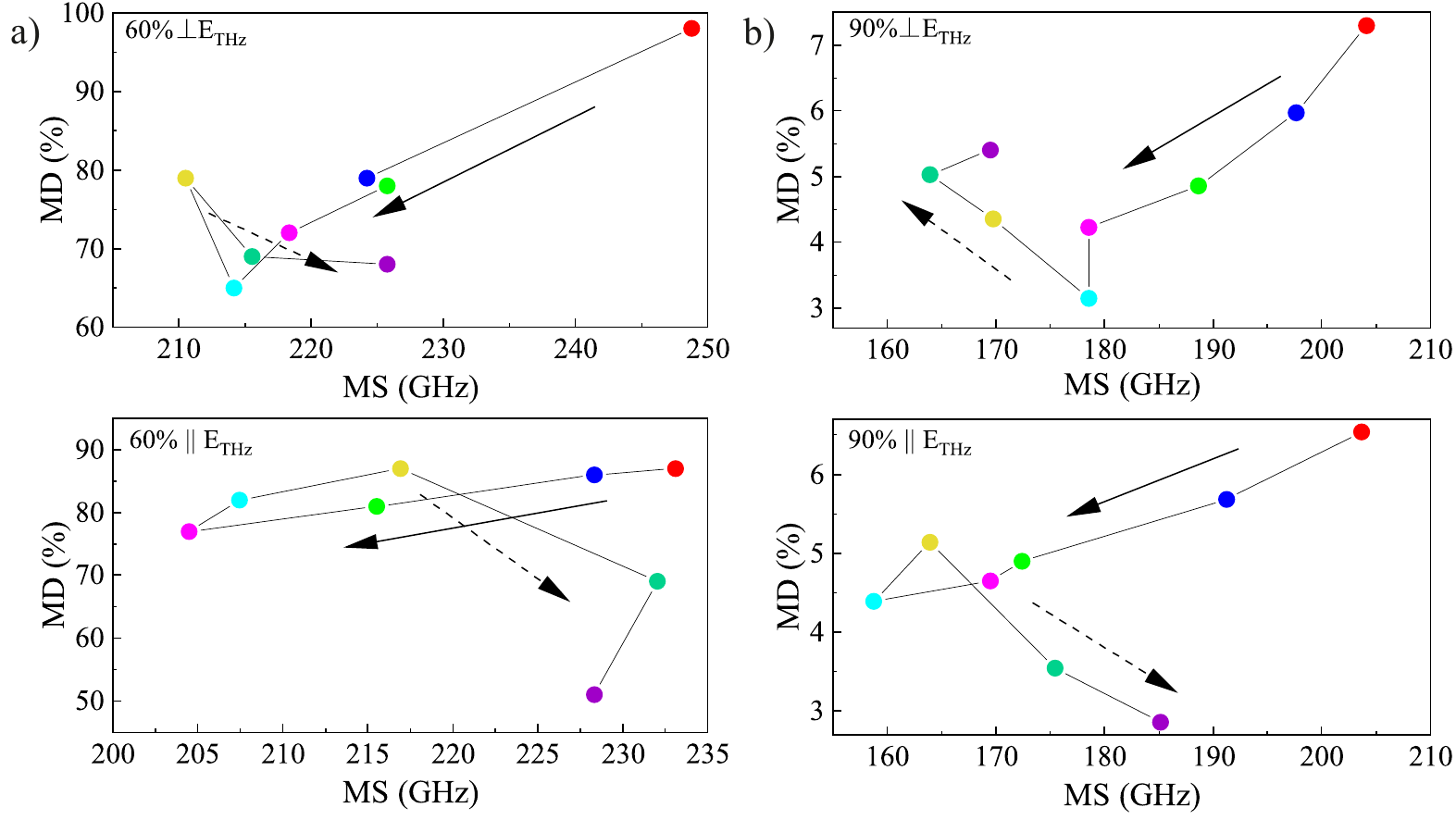}}
\end{center}
\caption{\label{FIG: Merite} The frequency-averaged MD (at the
peak of the time-domain THz pulse) at zero pump-probe delay versus MS for 60\% (a) and 90\% (b) films at perpendicular and parallel orientation. The colors of the points correspond to the colors in Fig.\ref{FIG: Transient}a.}
\end{figure}

In this work, the influence of stretching on optically controlled SWCNT modulators is experimentally demonstrated for the first time. 
To compare the efficiency of modulation under the applied stretching that was previously reported, we plotted MD versus MS (Fig. \ref{FIG: Merite}\marginnote{f5}). First of all, our modulator shows unprecedented almost 100\% MD which is accompanied by relatively high MS of almost 250 GHz. From one point of view, the tunability can be achived not only optically but also by stretching. Simultaneously, this information can be used to predict the performance of stretchable modulators based on SWCNT. Overall, the deterioration of the MD and MS was observed over the whole stretching circle. The MD exhibit a broad range of values reaching its minimum of 50\% in a parallel direction. A similar situation happens with MS which reached its minimum value of 205 GHz. We also compared the performance of thicker films with thinner 90\% SWCNT film. Full data are available in Fig. \ref{FIG: 90Transient} of SI. The obtained MD is significantly lower, varying from around 3\% to 7\%. We also found out that the MD of both films being divided on the thickness of a layer demonstrates almost the same MD, which is a sign of the straight dependence of MD on the film’s transparency (and, consequently, thickness). It is seen that MD lessens significantly with stretching in a forward direction up to 40\% of its initial length. The backward contraction reveals that MD before and after stretching is different, which reflects irreversible properties of SWCNTs photoconductivity.

\section{Conclusions}
We have demonstrated the influence of stretching on THz photoconductivity and performance of optically controlled THz modulators based on macro-scale SWCNT films. The high tunability of the modulators parameters over the full stretching circle was presented for SWCNT films on a stretchable substrate. The MD is continuously tuned from 100 to 70\% under stretching from 0 to 40\%. The MS is changing from 250 to 205 GHz. The physical mechanism of such behaviour is discussed in terms of the change of photoconductivity due to morphology change. A flexible broadband terahertz modulator based on a strain-sensitive SWCNT film is reported. As a conclusion, we would like to point out that the optical control of the proposed modulator allows to reach ultrafast operation frequencies, while its mechanical control enables the possibility to set MD and MS, which are required for for the communication system working at particular speed.

\section*{Acknowledgements}
M.G.B. and M.I.P. acknowledge Russian Science Foundation (RSF) Project No. 21-79-10097 (experimental OPTP measurements) and Ministry of Science and Higher Education of the Russian Federation No. 0714-2020-0002 (anisotropy factor measurements for SWCNT films).Also, M.G.B. acknowledge RSF Project No. 22-72-10033 (investigation of the stretchable modulators efficiency).  B.P.G. and A.G.N. acknowledge RSF Project No. 21-72-20050 (sample preparation). The authors would like to thank the Warwick Centre for Ultrafast Spectroscopy (University of Warwick, UK) for access to the OPTP spectrometer used.



\section*{Declaration of competing interest}
The authors declare that they have no known competing
financial interests or personal relationships that could have
appeared to influence the work reported in this paper.

\bibliography{mybibfile}

\clearpage

\beginsupplement

\begin{center}
\section*{{\LARGE{Supporting Information for:}}\\Ultrafast opto-mechanical terahertz modulators based on stretchable carbon
nanotube thin films }
\end{center}


\section*{Anisotropy}
Here we report the data on the aforementioned parameter of anisotropy for the samples of both thicknesses. 

\begin{figure}[h!]
\begin{center}
 \subfloat{\includegraphics[width=1\textwidth]{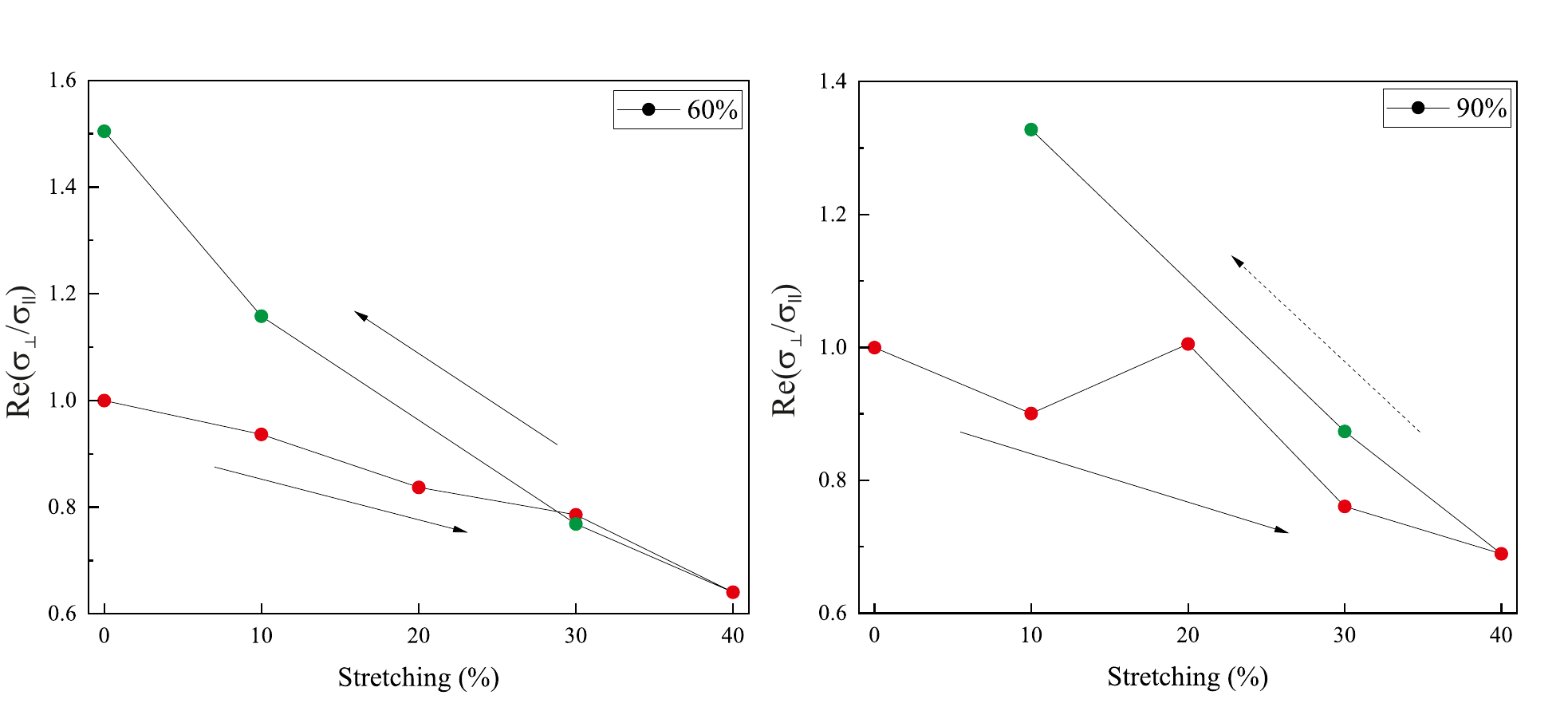}}
\end{center}
\caption{\label{FIG: SI Anisotropy} The anisotropy factor for SWCNT films with a) T=60$\%$ and b) T=90$\%$.}
\end{figure}
\clearpage
\section*{Frequency-dependent THz photoconductivity}

\begin{figure}[h!]
\begin{center}
 \subfloat{\includegraphics[width=1\textwidth]{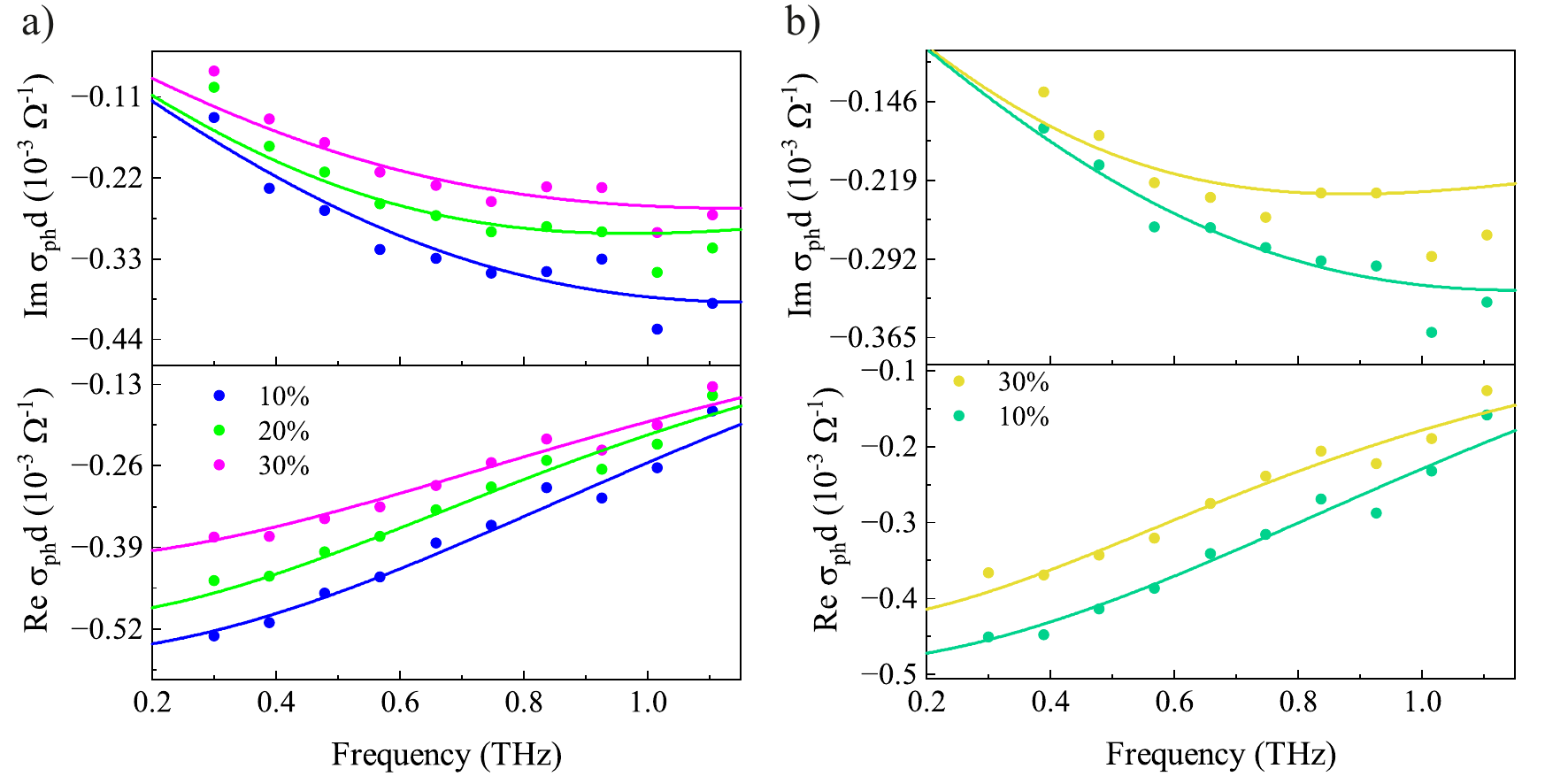}}
\end{center}
\caption{\label{FIG: SI Spectra} The film (T=60$\%$) sheet photoconductivity at 2 ps after photoexcitation of stretched SWCNT film a), and compressed film b) in a perpendicular direction.}
\end{figure}

\begin{figure}
\begin{center}
 \subfloat{\includegraphics[width=1\textwidth]{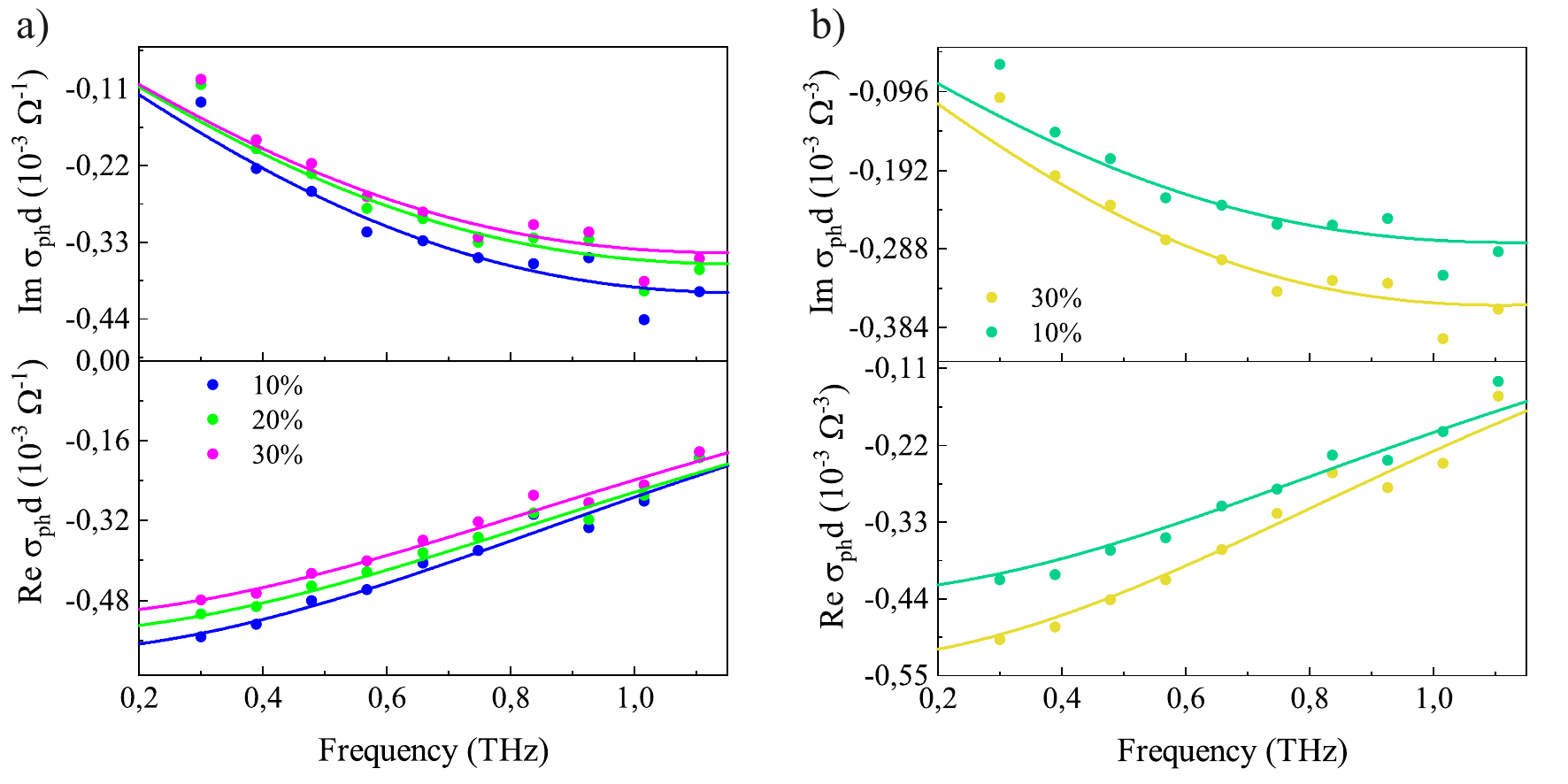}}
\end{center}
\caption{\label{FIG: SI Spectra1} The film (T=60$\%$) sheet photoconductivity at 2 ps after photoexcitation of stretched SWCNT film a), and compressed film b) in a parallel direction.}
\end{figure}

\begin{figure}
\begin{center}
 \subfloat{\includegraphics[width=1\textwidth]{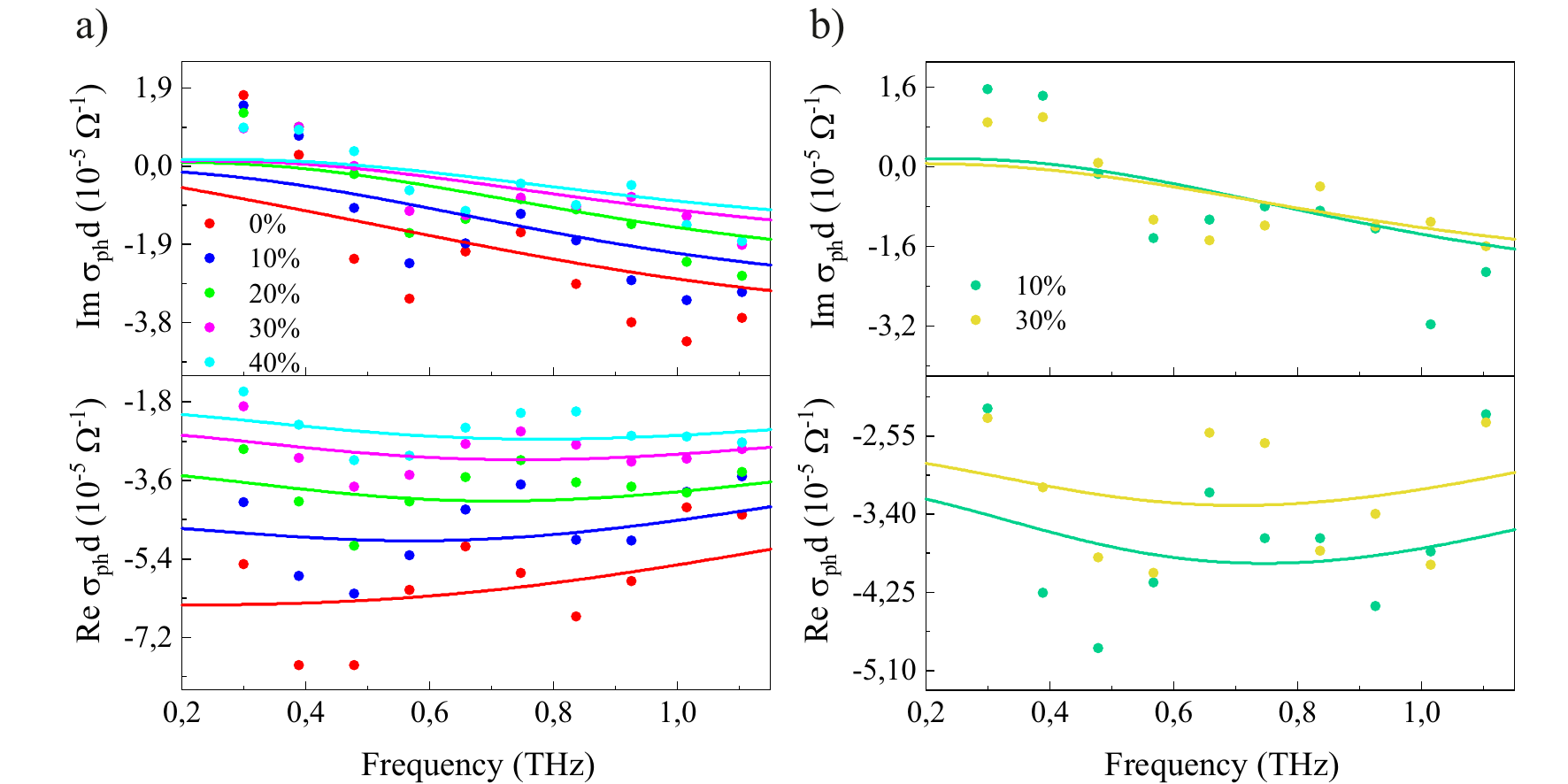}}
\end{center}
\caption{\label{FIG: SI Spectra2} The film (T=90$\%$) sheet photoconductivity at 2 ps after photoexcitation of stretched SWCNT film a), and compressed film b) in a perpendicular direction.}
\end{figure}

\begin{figure}
\begin{center}
 \subfloat{\includegraphics[width=1\textwidth]{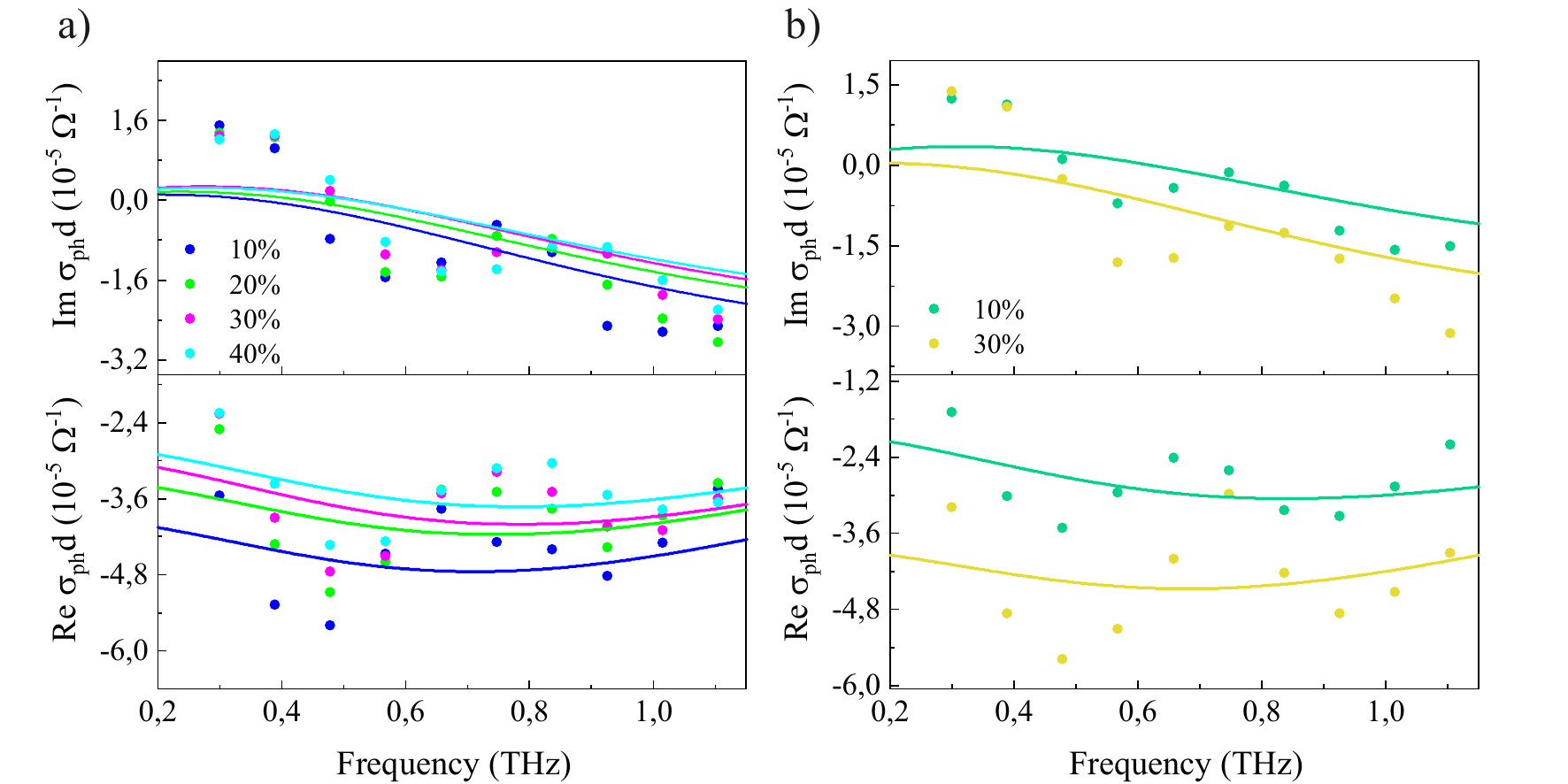}}
\end{center}
\caption{\label{FIG: SI Spectra3} The film (T=90$\%$) sheet photoconductivity at 2 ps after photoexcitation of stretched SWCNT film a), and compressed film b) in a parallel direction.}
\end{figure}

\begin{figure}[h!]
\begin{center}
 \subfloat{\includegraphics[width=1\textwidth]{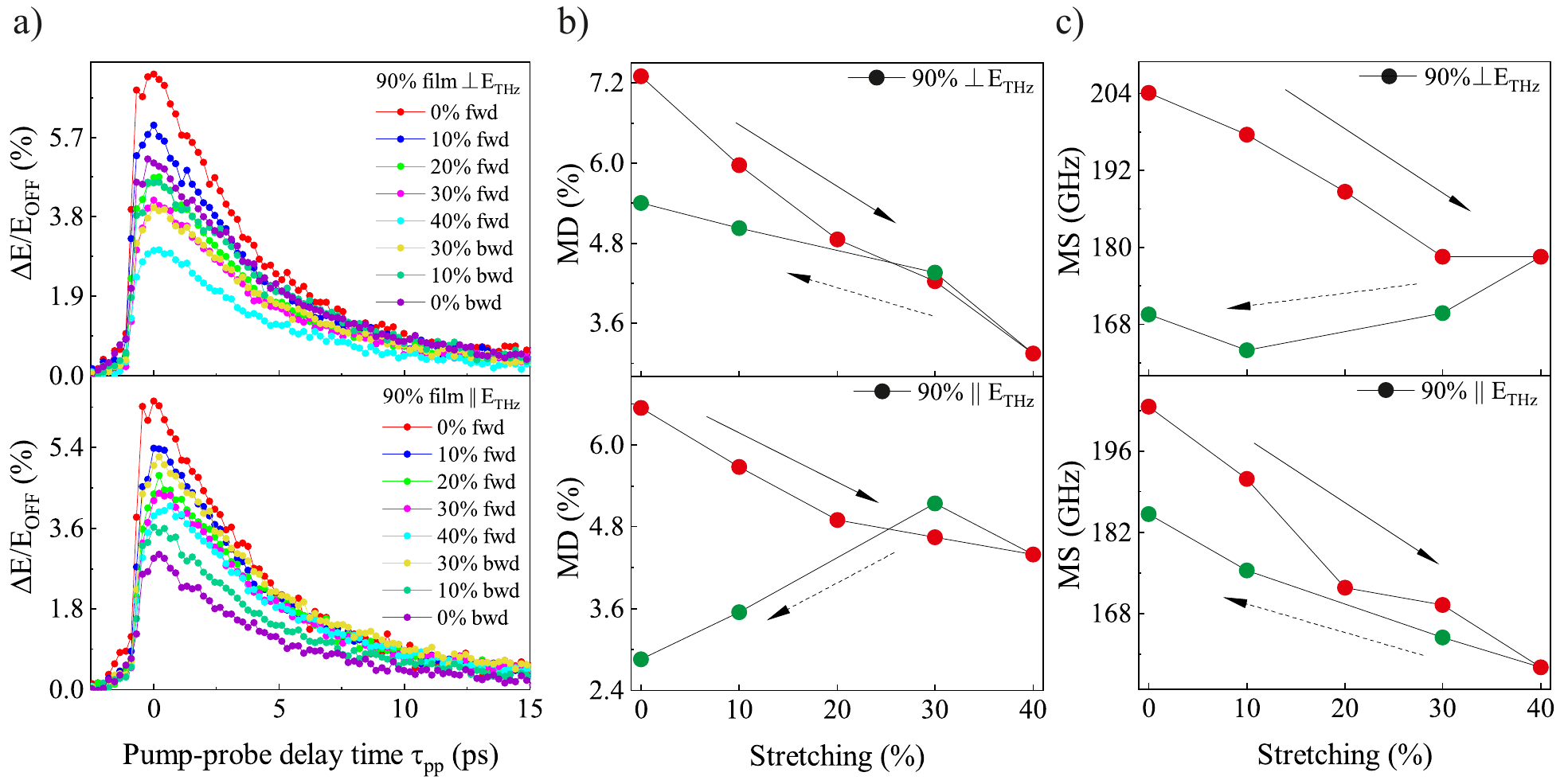}}
\end{center}
\caption{\label{FIG: 90Transient} (a) Modulation depth in transmission as a function of pump-probe delay time when the elongation is perpendicular (top) and parallel (bottom) to the THz probe beam. In transmission mode MD (b) and MS (c), for 90\% sample at stretching varying from 0 to 40\% and compression from 40 to 0\%. The arrows indicate the direction of stretching and the following compression.}
\end{figure}

The photoinduced frequency-domain spectra of the film with different strains were obtained in Fig. \ref{FIG: SI Spectra} -- \ref{FIG: SI Spectra3}. 

\clearpage
\section*{Data for 90\% film}

\end{document}